\documentclass[prd,twocolumn,showpacs,reprint,preprintnumbers,nofootinbib]{revtex4-1}

\usepackage[font=small,labelfont=bf,justification=centerlast]{caption}
\usepackage{array,multirow,makecell}
\usepackage{subcaption}
\usepackage{slashed}
\usepackage{hypernat}

\usepackage{graphicx}
\usepackage[nointegrals]{wasysym}
\usepackage{rotating}
\usepackage{color}
\usepackage{epsfig}
\usepackage{bm}
\usepackage{hyperref}
\usepackage{url}
\usepackage{slashed}
\usepackage{amsmath}
\usepackage{amssymb}

\usepackage{geometry}
\newcommand{\be}{\begin{equation}}
\newcommand{\ee}{\end{equation}}
\newcommand{\bea}{\begin{eqnarray}}
\newcommand{\eea}{\end{eqnarray}}

        \newcommand*{\refeq}[1]{Eq.~(\ref{#1})}

\def\lg{{\mathchoice{~\raise.58ex\hbox{$<$}\mkern-14.8mu\lower.52ex\hbox{$>$}~}
                    {~\raise.58ex\hbox{$<$}\mkern-14.8mu\lower.52ex\hbox{$>$}~}
                    {\raise.59ex\hbox{{$\scriptscriptstyle <$}}\mkern-12.8mu%
                     \lower.01ex\hbox{{$\scriptscriptstyle >$}}}   {}   }}
\def\gl{{\mathchoice{~\raise.58ex\hbox{$>$}\mkern-12.8mu\lower.52ex\hbox{$<$}~}
                    {~\raise.58ex\hbox{$>$}\mkern-12.8mu\lower.52ex\hbox{$<$}~}
                    {\raise.62ex\hbox{{$\scriptscriptstyle >$}}\mkern-12.0mu%
                     \lower.05ex\hbox{{$\scriptscriptstyle <$}}}  {}    }}

\begin{document}

\title{Self-interacting dark matter from late decays and the $H_0$ tension
}

\author{Andrzej Hryczuk}
\email{andrzej.hryczuk@ncbj.gov.pl}
\author{Krzysztof~Jod\l{}owski }
\email{krzysztof.jodlowski@ncbj.gov.pl}
\affiliation{National Centre for Nuclear Research, Pasteura 7, 02-093 Warsaw, Poland}

\date{\today}

\begin{abstract}
We study a dark matter production mechanism based on decays of a messenger WIMP-like state into a pair of dark matter particles that are self-interacting via exchange of a light mediator. Its distinctive thermal history allows the mediator to be stable and therefore avoid strong limits from the cosmic microwave background and indirect detection. A natural by-product of this mechanism is a possibility of a late time, {\it i.e.}, after recombination, transition to subdominant dark radiation component through three-body and one-loop decays to states containing the light mediator. We examine to what extent such a process  can help  to alleviate the $H_0$ tension. Additionally, the mechanism can provide a natural way of constructing dark matter models with ultrastrong self-interactions that may positively affect the supermassive black hole formation rate. We provide a simple realization of the mechanism in a Higgs portal dark matter model and find a significant region of the parameter space that leads to a mild relaxation of the Hubble tension while simultaneously having the potential of addressing small-scale structure problems of $\Lambda$CDM.
\end{abstract}

\maketitle

\section{Introduction}

The standard $\Lambda$CDM cosmological model incorporates dark matter (DM) in the simplest way possible, {\it i.e.,} a noninteracting cold matter component with constant equation of state throughout its cosmological evolution. A scenario of this type is not only simple and remarkably successful in explaining the Universe at large scales but also well motivated in many theories beyond the Standard Model (SM) of particle physics. However, the shortcomings of $\Lambda$CDM at small scales, {\it e.g.}, the diversity \cite{Oman:2015xda,Kamada:2016euw}, too big to fail \cite{Boylan_Kolchin_2011}, missing satellites  \cite{Moore:1999nt,Klypin:1999uc,Fattahi:2016nld} and core-cusp \cite{deBlok:1997zlw,Oh:2010ea,Walker:2011zu} problems, as well as tensions between parameters inferred from local and global cosmological measurements, most notably the Hubble parameter $H_0$ \cite{Aghanim:2018eyx,Riess:2016jrr,Riess:2019cxk} (see, {\it e.g.,} \cite{Knox:2019rjx} for a review), may be viewed as a hint that the CDM paradigm is in fact too simple. Indeed, it is well known that at least some of the small scale problems  can be simultaneously addressed if DM possesses significant self-interactions preferably with velocity-dependent cross section (see, {\it e.g.,}~\cite{Tulin:2017ara} for a review). Additionally, varying equation of state, {\it e.g.,} due to late time conversion of a small fraction of DM into radiation, has been shown to have the potential for reducing the $H_0$ tension~\cite{Blackadder:2015uta,Vattis:2019efj} (but see also~\cite{Haridasu:2020xaa,Clark:2020miy}; for a related, but different approach see~\cite{Gu:2020ozv}).

It is an intriguing question if both small scale problems and  $\Lambda$CDM tensions can be simultaneously resolved through a modification of only the DM component. This point has been addressed in thermally produced self-interacting dark matter models featuring strong Sommerfeld enhancement in \cite{Binder:2017lkj,Bringmann:2018jpr}, where it has been demonstrated that  late time annihilations can indeed be efficient enough to sufficiently modify the cosmological evolution.

However, models predicting thermally produced DM self-interacting via light mediator often run into problems with observations (see {\it e.g.},~\cite{Bringmann:2016din}). The DM annihilation to the mediator pair  is greatly enhanced during the recombination epoch by the Sommerfeld effect \cite{Hisano:2004ds,ArkaniHamed:2008qn} leading to too large energy injection into the plasma, if the mediator decays to visible states. On the other hand, for stable light mediators the overclosure bound is greatly constraining due to their large thermal population.

Several possibilities of how to avoid such limits are known, {\it e.g.,} by having the mediator decay only to neutrinos or dark radiation (subject to much weaker bounds) or by assuming that the dark sector (DS) is effectively secluded and has much lower temperature than the one of the photon bath. In this paper we propose to utilize a mechanism for DM production akin to the one used in the superWIMP scenario \cite{Feng:2003xh} and show that it introduces alternative way of constructing models with velocity dependent self-interactions. In such a setting the DM component arises from decays of an intermediate weakly interacting massive state, which in turn is thermally produced via the usual freeze-out process. This production mode allows the mediator of the interactions in the DS to be absolutely stable while at the same time not overclosing the Universe.  Additionally, if the decays of the WIMP-like particle happen at very late times, it is exactly the framework needed for the conversion of  dark matter to radiation that might help alleviate the $H_0$ tension. What is more, in this mechanism it is quite natural to expect that only a small fraction of WIMPs decay into light mediators, as it is a higher order process compared to the tree-level decay to the DM particles.

This paper is organized as follows. In Sec.~\ref{sec:model} we introduce the mechanism and the example from a generic class of Higgs portal models. Section~\ref{sec:pheno} describes the thermal history, lays out calculations of DM self interactions and late time decay impact on cosmology. In Sec.~\ref{sec:results} we show and discuss the results of the numerical analysis. Finally, we conclude in Sec.~\ref{sec:conclusions}.

\section{The mechanism}
\label{sec:model}

The main idea behind the production mechanism studied here is that, if the dark sector is populated by decays taking place late enough that it never reaches chemical equilibrium with the visible sector, then the light mediator is effectively absent from the plasma while still carrying a long range force between DM particles. Therefore, it can be absolutely stable and completely naturally evade all the limits from CMB observations and indirect detection.\footnote{Although, for simplicity, we will limit ourselves to stable mediators, we remark that introducing a small decay width provides a model which is still viable and with additional potential phenomenology and detection possibilities.}

\subsection{The SM-DS coupling through a portal }

\begin{figure}[t]
\centering
\includegraphics[scale=0.25]{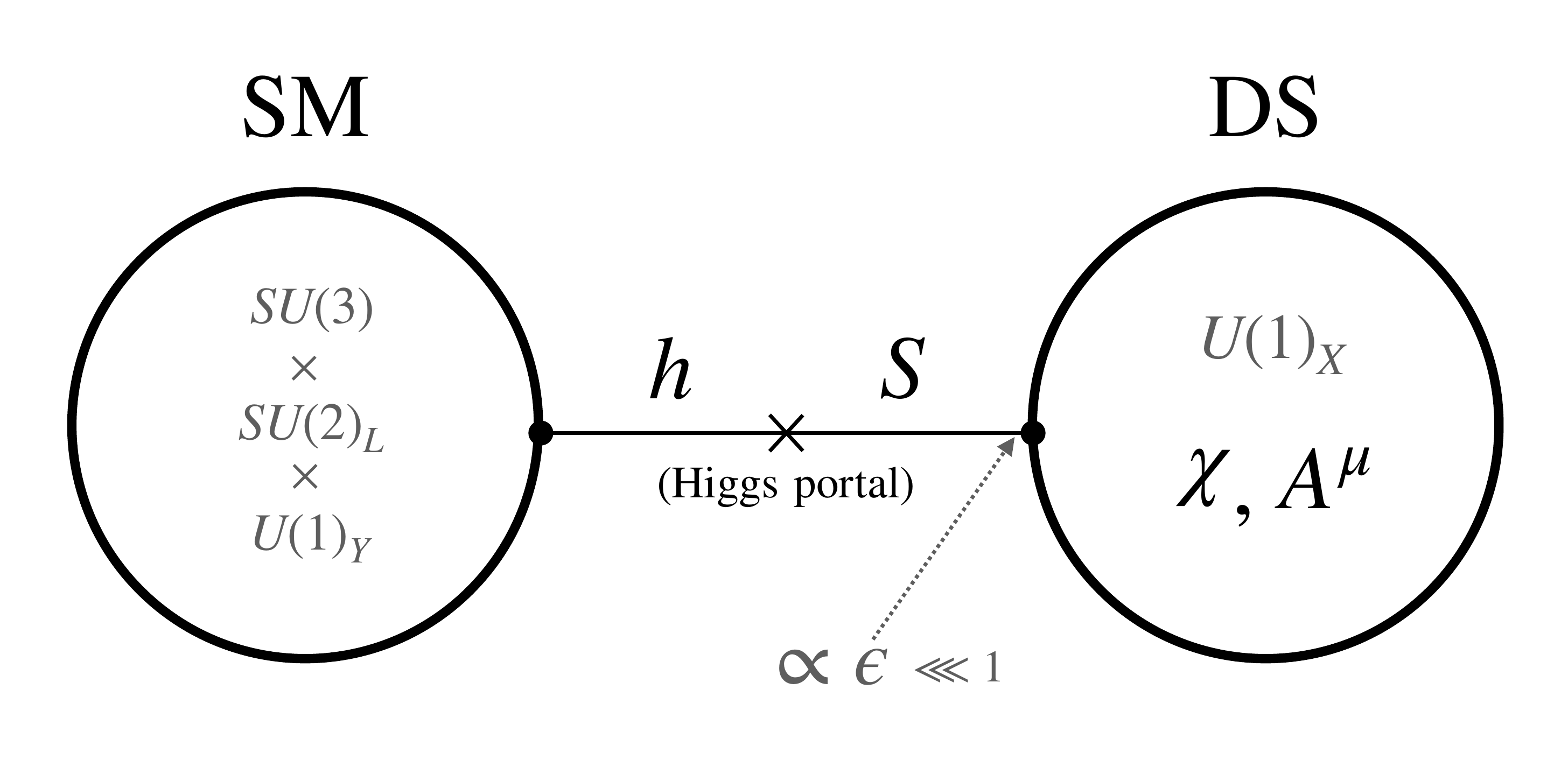}
\caption{The schematic picture of the setup. The visible SM sector is connected through a Higgs portal connector $S$ to the dark sector, where the latter is built up of a Dirac fermion $\chi$ charged under gauged $U(1)_X$ with massive gauge field $A^\mu$.}
\label{fig:model}
\end{figure}

A very generic framework naturally encompassing the above mechanism is the scenario when the dark sector is connected to the visible sector only through a weak portal. Here, for concreteness, let us concentrate on a so-called Higgs portal connection between the SM and the DS, which is one of the most simple and natural choices. Multitude of examples of such DM models can be found in the literature (see, {\it e.g.,} \cite{Arcadi:2019lka} for a review). The scenario is illustrated in Fig.~\ref{fig:model} where the connecting SM singlet scalar $S$ is assumed to mix weakly with the Higgs and also have a weak or very weak coupling to the states in the dark sector.\footnote{The simplest realization of such setup would assume only one state in the dark sector, which would be stable and provide the DM candidate. Such case, however, cannot accommodate any significant velocity dependent self-interactions between DM particles.}

A natural choice for $S$ is to be a pseudo-WIMP, {\it i.e.,} particle undergoing thermal freeze-out with near-stability guaranteed by imposed spontaneously or explicitly broken $Z_2$ symmetry $S \leftrightarrow -S$.

The perspective of DM portal framework highlights an alternative angle on the studied mechanism: it can be viewed as an extension of the usual Higgs portal freeze-out or freeze-in models to even weaker couplings to the dark sector. Indeed, parametrizing the breaking by a small parameter $\epsilon$, one can quite generally  distinguish four regimes:

\begin{enumerate}
    \item[0)] weak $\lesssim \epsilon$: the DS reaches chemical equilibrium with the SM independently of the reheating details leading to a thermal population of the dark matter and light mediator - one recovers \textit{usual thermal self-interacting model subject to strong limits}

    \item[A)]  very weak $\lesssim\epsilon \lesssim$ weak: the DS is produced through decay of $S$ and never reaches chemical equilibrium with the SM; the light interaction mediator can be stable and avoid overclosure and CMB limits; \textit{viable regime for self-interacting DM}

    \item[B)]  ultra weak $\lesssim \epsilon \lesssim$ very weak: the same as A, but leading to $S$ having lifetime on cosmological scales; \textit{regime for self-interacting DM with an impact on the $H_0$ tension}

    \item[C)]  $\epsilon \lesssim$ ultra weak: $S$ is quasistable with onset of its decays reaching times of order of Gyr; one ends up with two component DM with only a fraction being self-interacting which can play a role of ultrastrong self-interacting dark matter (uSIDM)~\cite{Pollack:2014rja}; \textit{regime potentially addressing the $H_0$ tension and providing an uSIDM candidate}.
\end{enumerate}

Regimes 0 and A point to the $Z_2$ breaking at relatively low  energy scales, not much larger than the DM particle mass. Smaller values of $\epsilon$ leading to scenarios B and C naturally emerge when the breaking comes from some new physics at a very high scale, {\it e.g.,} GUT or even Planck scale.

\subsection{Toy model example}

For concreteness, let us consider a dark sector comprised of a Dirac fermion $\chi$ charged under new gauged $U(1)_X$ broken spontaneously at some higher scale resulting in massive vector $A^\mu$.\footnote{The symmetry breaking of $U(1)_X$ can, but does not have to be, related to the breaking of the $Z_2$.} This choice is not crucial in what follows, but exemplifies a very simple and natural realization within a renormalizable model.

The dark sector part of the Lagrangian after the $U(1)_X$ breaking reads
\begin{eqnarray}
\mathcal{L}^{\rm DS} =\ & &
\bar{\chi}( i \gamma_\mu \partial^\mu-m_{\chi})\chi + \frac{1}{2}m_A^2 A_\mu A^\mu \nonumber\\
&&  + igA^\mu \bar\chi \gamma_\mu \chi + \epsilon \, S \bar{\chi} \chi
\label{eq:LDS}
\end{eqnarray}
while the connection with the visible sector is given by the portal
\begin{eqnarray}
\mathcal{L}^{\rm portal} =\ &&  \frac{1}{2} (\partial^\mu S )  (\partial_\mu S) + \frac{\mu_{S}^{2}}{2} \, S^2 + \frac{\lambda_3}{3!} \, S^3  + \frac{\lambda_4}{4!} \, S^4  \,  \nonumber\\
&& + \epsilon\,\mu_{HS} S \, H^{\dagger} H    +   \lambda_{HS} \, S^2 H^{\dagger} H \, ,
\label{eq:Lportal}
\end{eqnarray}
where $H$ denotes the SM Higgs boson doublet and in the trilinear term we explicitly pulled out the $\epsilon$ factor to emphasize that this term is allowed only due to $Z_2$ breaking. This is a crucial observation because it ensures that $S$ decays predominantly to DS states, if only $\mu_{HS}$ is small enough or $S$ light enough that the resulting branching ratio to the SM particles is strongly suppressed compared to BR($S\to \bar\chi\chi$).
Phenomenologically interesting interactions are given in the second lines of both Eq.~\eqref{eq:LDS} and~\eqref{eq:Lportal}.

\section{Phenomenology}
\label{sec:pheno}

Having introduced the framework and defined concrete realization we describe in this section the main properties of such a scenario.

\subsection{Thermal history}

\begin{figure*}[ht]
\begin{subfigure}{1\textwidth}
\centering
\includegraphics[width=.95\textwidth]{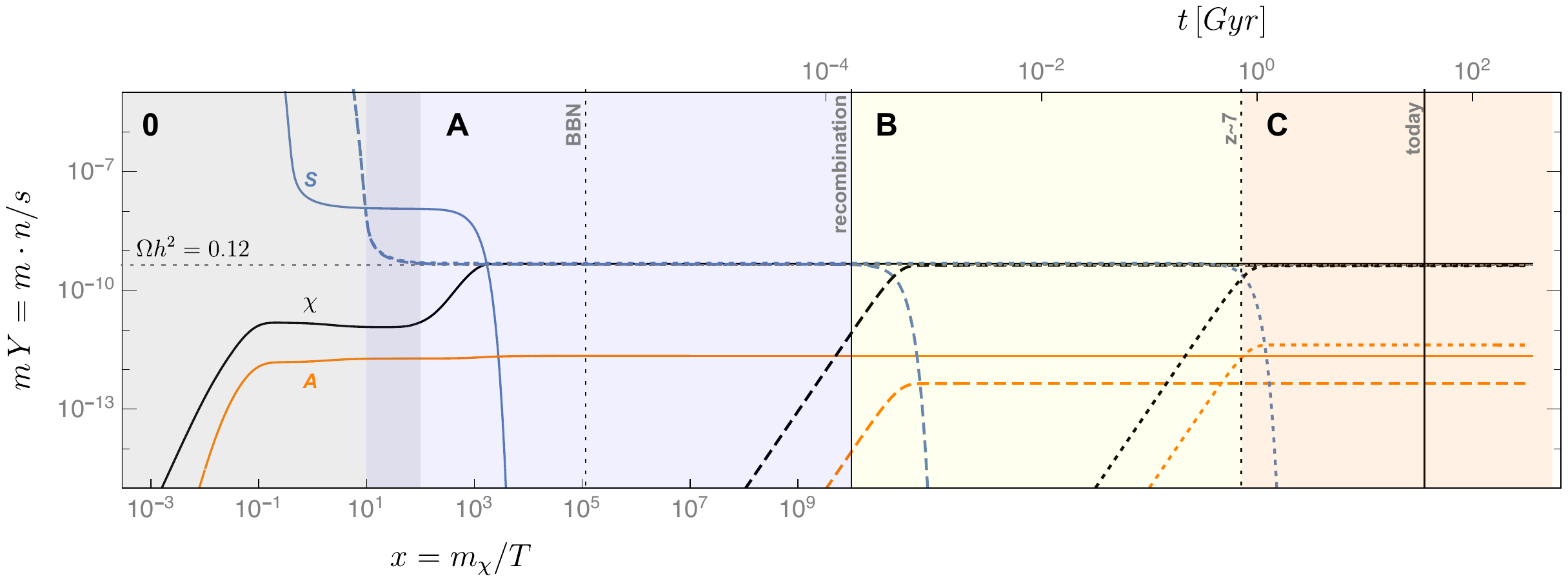}
\end{subfigure}
\caption{The illustration of the thermal history of $S$ (blue), $\chi$ (black) and $A^\mu$ (orange) with example parameter choices leading to early (regime A, solid lines), late (regime B, dashed) and very late (regime C, dotted) decays of $S$. The borders of the regimes are indicative and not sharply defined. In particular, the redshift $z\sim 7$ line corresponds to times of oldest observed quasars with SMBHs - see text and Sec.~\ref{sec:uSIDM} for details.
}
\label{fig:Y}
\end{figure*}

The underlying assumption in the discussion of the thermal history of $\chi$ is the one of the freeze-in models, {\it i.e.,} that only SM sector is populated during reheating, while the dark sector has negligible initial number and energy densities.

The connector $S$ undergoes usual WIMP-like evolution where it thermalizes with the SM plasma at the early times  due to mixing and, typically more importantly, the quartic $ \lambda_{HS}$ coupling. When its annihilation rate drops below the Hubble rate it goes through the freeze-out process. At later times, possibly even after recombination, it decays via $S\rightarrow \bar\chi\chi$ and also, by construction, subdominantly to SM through the Higgs mixing. In Fig.~\ref{fig:Y} an illustration of example evolution of mass densities of $S$, $\chi$ and $A^\mu$ is shown
for decay regimes A (solid lines), B (dashed) and C (dotted). In all the cases the $\chi$ and $A^\mu$ undergo a freeze-in type production, which is very inefficient due to smallness of the coupling to $S$. It follows that their number densities are extremely small until the onset of $S$ decay.

The transitions between the regimes are only indicative and not sharply defined. In particular, the chosen redshift $z\sim 7$ line separating cases B and C corresponds to times of oldest observed quasars with supermassive black holes (SMBHs) \cite{Mortlock_2011,DeRosa:2013iia,Banados:2017unc}. Decays of $S$ around that time can impact the formation rate of the SMBHs, see Sec.~\ref{sec:uSIDM}.
The onset of $S$ decays can also happen later until and beyond the present day, meaning that regime C extends to cover all the possible lifetimes of $S$.

As can be seen in Fig.~\ref{fig:Y} in case A the connector $S$ typically needs to chemically decouple with larger number density than would give the correct thermal abundance, since during the decay some of its energy is transferred to the kinetic energy of the $\chi$, which gets redshifted. Note also that annihilation of $\chi\bar\chi \to AA$ can have some effect, even if the number densities do not reach equilibrium values, as seen in the small drop of $\chi$ density at early times. For later decays in case B and C the $\chi$ particles need to be produced with very small kinetic energy, as discussed in Sec.~\ref{sec:structures} below, otherwise will negatively affect the structure formation. It follows that $S$ needs to have the number density just a bit over the observed one which is then nearly completely transferred to the DM.

\subsection{Dark matter self-interactions}

In calculating the strength of the elastic scattering between two DM particles at present day velocities $v\sim10^{-3}$ we  follow standard numerical procedure of solving Schr\"{o}dinger equation described in~\cite{Buckley:2009in,Tulin:2017ara}. We use natural units $c=\hbar=1$.

The relevant quantity with respect to self-interactions is transfer cross section which is defined as a weighted average of the differential cross section with respect to the fractional longitudinal momentum transfer $(1-\cos \theta)$:
\begin{equation}
\begin{aligned} \sigma_{\mathrm{tr}} & \equiv \sigma := \int d \Omega(1-\cos \theta) \frac{d \sigma}{d \Omega} \\ &=\frac{4 \pi}{k^{2}} \sum_{\ell=0}^{\ell_{max}}\big[(2 \ell+1) \sin ^{2} \delta_{\ell}-2(\ell+1) \sin \delta_{\ell} \times \\ & \times  \sin \delta_{\ell+1} \cos \left(\delta_{\ell+1}-\delta_{\ell}\right)\big].
\end{aligned}
\label{eq:transfer_sigma}
\end{equation}

The differential cross section is given by series expansion into Legendre polynomials corresponding to orthogonal partial waves:
\begin{equation}
\frac{d \sigma}{d \Omega}=\frac{1}{k^{2}}\left|\sum_{\ell=0}^{\ell_{max}}(2 \ell+1) e^{i \delta_{\ell}} P_{\ell}(\cos \theta) \sin \delta_{\ell}\right|^{2}.
\end{equation}

The phase shift $\delta_\ell$ for a partial wave $\ell$ is obtained by solving Schr\"{o}dinger equation for the radial wave function $R_\ell(r)$, which describes reduced $\chi$-$\chi$ system, given by
\begin{equation}
\frac{1}{r^{2}} \frac{d}{d r}\left(r^{2} \frac{d R_{\ell}}{d r}\right)+\left(k^{2}-\frac{\ell(\ell+1)}{r^{2}}-2 \mu V(r)\right) R_{\ell}=0,
\label{eq:schrodinger_eq}
\end{equation}
where $v$ is relative velocity of $\chi$'s, $\mu=m_\chi/2$ is reduced mass of the system and $k=\mu v$. Potential term comes from the gauge interactions in Eq.~\eqref{eq:LDS}. Multiple exchanges of $A^\mu$ coupled to $\chi$ with coupling strength $\alpha = g^2/(4\pi)$, result in a Yukawa-type potential:
$$
V(r)=\pm\frac{\alpha}{r} e^{-m_{A} r}.
$$

Since we took $A^\mu$ to be a vector, the interactions are attractive ($-$) for $\chi\bar{\chi}$ scattering and repulsive ($+$) for $\chi\chi$ or $\bar{\chi}\bar{\chi}$ scattering. The interaction cross section is then taken as the average of repulsive and attractive interactions.

Far away from the Yukawa potential range \refeq{eq:schrodinger_eq} has well known solution in terms of spherical Bessel functions $j_\ell(r)$ and $n_\ell(r)$ (for definitions and properties of spherical Bessel functions see, {\it e.g.}, Sec.~10.47 in~\cite{10.5555/1830479}):
\begin{equation}
\lim _{r \rightarrow \infty} R_{\ell}(r) \propto \cos \delta_{\ell} j_{\ell}(k r)-\sin \delta_{\ell} n_{\ell}(k r)
\end{equation}

Therefore, one needs to numerically solve \refeq{eq:schrodinger_eq} for $a \leq r \leq b$ and match numerical solution at $b$ to the analytic one. We use Numerov method~\cite{numerov:1,numerov:2} which is fourth-order linear method in the step size $h=(b-a)/n$, where $n$ is number of points in the grid.
Limiting points $a$ and $b$ are determined by demanding that at $a$ \refeq{eq:schrodinger_eq} is dominated by the centrifugal term, which means $a \ll \frac 1 {m_A}, \frac {\ell+1}{\mu v}$.
The upper bound, $b$, is determined by demanding that the potential term is much smaller than the kinetic term:
$\frac{\alpha}{b} e^{-m_{A} b} \ll \frac{\mu v^2}{2} $.

The resulting phase shift is determined by matching the numerical solution with asymptotic one at the endpoint of the grid~\cite{Schmid:1979pwy}:
\begin{equation}
\tan \left(\delta_{\ell}\right)=\frac{ j_{\ell}(k(b-h)) R_{\ell}(b)- j_{\ell}(k b) R_{\ell}(b-h)}{n_{\ell}(k(b-h)) R_{\ell}(b)- n_{\ell}(k b) R_{\ell}(b-h)},
\end{equation}
where $R_{\ell}$ is a wave function obtained numerically and $j_{\ell}$, $n_{\ell}$ are spherical Bessel functions.

We calculate phase shifts until convergence of \refeq{eq:transfer_sigma} where we consider $\sigma$ to be converged if successive values obtained for $\ell_{max}$ and $\ell_{max}\to \ell_{max}+1$ differ by less than 0.1\%.

The numerical solution is strictly needed only in the resonant regime, which occurs when $\frac{\alpha m_\chi}{m_A} \gtrsim 1$. In other regions of parameter space one can use analytic formulas to speed up the numerical scan. These can be obtained either from perturbative expansion in $\alpha$ (Born regime~\cite{Feng:2009hw}; applicable when $\frac{\alpha m_\chi}{m_A} \ll 1$) or from classical calculations of charged particles moving in plasma (classical regime~\cite{Feng:2009hw,Khrapak:2003kjw,Khrapak:2014xqa,Tulin:2012wi}; applicable when  $\frac{v m_\chi}{m_A} \gg 1$). We find agreement between numerical results and analytic formulae whenever they are applicable.

Both the coupling $g$ and light mediator mass $m_A$ governing the scattering cross section are free, essentially unconstrained parameters of the model. It follows that a very wide range of possible self-interaction strengths can be obtained. Two regions are of particular phenomenological interest, on which we will focus:
\begin{itemize}
    \item The first is when $\sigma/m_\chi \in (10^{-1},10^1)\,$g/cm$^2$ leading to momentum transfer rates in the correct ballpark to address the small-scale structure problems of $\Lambda$CDM. Theories giving rise to cross sections in this range are often referred to as the strongly interacting dark matter (SIDM) models.
    \item The second is the so-called ultra SIDM (or uSIDM) regime with $\sigma/m_\chi \gtrsim 10^3$g/cm$^2$  which could resolve the puzzle of supermassive black holes formation. One possible solution is that a small uSIDM component can, through a gravothermal collapse, form an initial seed which is what is needed for accelerating growth rate of SMBHs at their early stages of evolution~\cite{Pollack:2014rja,Choquette:2018lvq}.
\end{itemize}

\subsection{Late time $S$ decay}

Due to the breaking of the stabilizing $Z_2$ symmetry, the $S$ decays both to DS and SM states. We will assume that the latter are negligible compared to the former, which is the case if only the trilinear coupling $\mu_{HS}$ is small enough leading to small mixing with the Higgs. At tree-level the only decay is then $S \to  \bar\chi \chi$ with width taking the form:
\begin{eqnarray}
&&\Gamma_{S \to \bar{\chi} \chi}  = \frac{\epsilon^2}{8 \pi} \frac{(m_{S}^{2}- 4m_{\chi}^{2} )^{3/2} }{m_{S}^{2}} \\
&&\approx 5.3\times 10^4 \! \left(\frac{m_S}{1 {\rm GeV}} \right) \! \left(\frac{\epsilon}{10^{-16}} \right)^2 \! \left(\frac{\delta }{10^{-4}} \right)^{3/2} \! {\rm km/s/Mpc}  \nonumber
\label{eq:GammaS}
\end{eqnarray}
where
\begin{equation}
\label{eq:mass_splitting}
\delta \equiv 1-\frac{2m_\chi}{m_S}
\end{equation}
is the parameter governing the mass splitting and we introduced exemplary  parameter values that lead to late decays.

However, at higher order the three-body $S \to \bar{\chi} \chi A$ and loop decay $S \to A A$ are present and parametrically $\Gamma_{S \to \bar{\chi} \chi A}/\Gamma_{S \to \bar{\chi} \chi} \sim g^2$ and $\Gamma_{S \to  A A}/\Gamma_{S \to \bar{\chi} \chi} \sim g^4$
where the former is also potentially significantly affected by the available phase space, especially if $\delta \ll 1$. One can see that $S$ decay \textit{naturally results in few \% of energy being transferred to radiation} and therefore one obtains a complete one-component DM model with the property desired for alleviating the $H_0$ tension.

In more detail, final decay products will be either nonrelativistic (in tree decay $S \to  \bar\chi \chi$, act as dark matter), relativistic (in loop $S \to A A$, act as dark radiation) or mixed (in three body $S \to \bar{\chi} \chi A$). In the latter case we adopt a prescription that $\chi$ will always act as matter (very good approximation as long as $\delta$ is small, as assumed), while $A^\mu$ will be counted as matter if its kinetic $E_A < m_A$, otherwise as radiation.\footnote{We have checked that adopting different definition of separating relativistic and nonrelativistic regions of the phase space has only slight effect on our results.}

The differential three-body decay rate reads
\begin{equation}
\frac{d^2\Gamma_{S \to \bar{\chi} \chi A}}{dE_A dE_\chi}  = \frac{|\mathcal{M}_{S \to \bar{\chi} \chi A}|^2}{64 \pi^3 m_S},
\label{eq:GammaSChiChiPhi}
\end{equation}
where the amplitude $\mathcal{M}_{S \to \bar{\chi} \chi A}$ is given by:
\begin{eqnarray}
\mathcal{M}_{S \to \bar{\chi} \chi A} &=& \epsilon \, g \, \epsilon^{\mu *}_r(p_1) \bar{\chi}(p_3,m_\chi) \biggl(\frac{1}{\slashed{p}_1 +\slashed{p}_3-m_\chi}\,  \nonumber\\
& - & \frac{1}{\slashed{p}_1+\slashed{p}_2+m_\chi}\biggr)\chi(p_2,m_\chi),
\end{eqnarray}
where $p_1$ is momentum of $A$ and $p_2$, $p_3$ are momenta of $\bar{\chi}$ and $\chi$, respectively, in the rest frame of $S$. $\epsilon^*_r(p_1)$ is polarization vector coming from external $A^\mu$.

Integrating over the whole kinematically allowed region, we get total $\Gamma_{S \to \bar{\chi} \chi A}$. However, to calculate the fraction of energy transferred to radiation we need to separate the region where $A^\mu$ is relativistic at decay. We will approximate this fraction by the quantity:
\begin{equation}
\text{F} = \frac{ \Gamma_{S\to AA} + \Delta \times \Gamma_{S\to \bar\chi\chi A}}{\Gamma_{S \to \bar\chi\chi} + \Gamma_{S\to AA} + \Gamma_{S\to \chi\chi A}} \, ,
 \label{eq:F}
\end{equation}
where
\begin{equation}
\Delta = \frac{1}{\Gamma_{S \to \bar{\chi} \chi A}}\int_{2 m_A}^{E_A^{max}} \int_{E_\chi^{min}}^{E_\chi^{max}}\frac{d^2\Gamma_{S \to \bar{\chi} \chi A}}{dE_\chi dE_A}dE_\chi dE_A
\end{equation}
is the fraction of the decay width resulting in $A^\mu$ having kinetic energy equal or larger to its mass.

The one loop decay $S\to AA$ is of a higher order in perturbation theory, but does not suffer from phase space suppression and transfers all the energy of $S$ to radiation. For calculations we used \texttt{Mathematica} packages \texttt{FeynCalc}~\cite{Mertig:1990an,Shtabovenko:2016sxi,Shtabovenko:2020gxv} and \texttt{Package-X}~\cite{Patel:2016fam} to symbolically calculate the amplitude and evaluate the numerical expressions:
\begin{equation}
\Gamma_{S\to AA}=\frac{|\mathcal{M}_{S\to AA}|^2}{(16\pi^2)^2}\frac{g^4 \epsilon^2\sqrt{m_S^2-4m_A^2}}{16\pi m_S^2}.
\label{eq:Sphiphi}
\end{equation}
The amplitude $\mathcal{M}_{S\to AA}$ is given by:
\begin{eqnarray}
\mathcal{M}_{S\to AA}\! &=& -12 m_\chi \big[-2B_0\left(m_S^2;m_\chi,m_\chi\right)\\
& + &\! 8C_{00}(m_A^2,m_S^2,m_A^2;m_\chi,m_\chi,m_\chi)\nonumber\\
&+ &\! (2m_A^2 \! - \!m_S^2)C_0 \! \left(m_A^2,m_S^2,m_A^2;m_\chi,m_\chi,m_\chi \right)\! \big],\nonumber
\label{eq:amp_Sphiphi}
\end{eqnarray}
where $B_0$ and $C_0$ are two and three-points Passarino--Veltman~\cite{Passarino:1978jh} scalar functions, respectively, while $C_{00}$ is coefficient of three-point tensor function proportional to the metric. We follow conventions of~\cite{Patel:2016fam}, where,
in particular, the $1/(16\pi^2)$ is factored-out in their expressions, hence it reappears in \refeq{eq:Sphiphi}.
Note that $B_0$ and $C_{00}$ are UV divergent, however their divergent parts actually cancel out in \refeq{eq:amp_Sphiphi}, which renders the whole expression finite.

Before concluding this subsection a comment is in order. If $m_S\approx 2m_\chi$, which as we discuss later is expected to be necessary not to spoil large structure formation, then the $\chi$s produced in $S$ decay will have small velocities. Since they interact via light mediator creating long range force, there can be a substantial threshold correction. If present, it would mainly result in a shift of the $\epsilon$ coupling which is not consequential for what follows. The reason is that such a threshold effect would appear in all three decay processes and while one would expect some change in their relative size the inclusion of this effect would be necessary only when high precision is called for and goes beyond the scope of our work.

\subsection{$H_0$ tension and structure formation}

In recent years, cosmological probes become increasingly more precise which further constraints alternatives to the standard $\Lambda$CDM model. One of the persistent tensions, which actually became more severe with more data, is determination of Hubble parameter. Early Universe observations such as CMB or baryonic acoustic oscillations (BAO) prefer significantly lower value $H_0\sim67\,$km/s/Mpc in comparison to the local Universe observations which determine $H_0 \sim 74\,$km/s/Mpc. The uncertainties of the measurements are $\sim\,$1--2\% and the resulting discrepancy reaches $\sim 4 \sigma$. Currently no universally accepted solution is known~\cite{Knox:2019rjx}, however it is believed that systematic errors in both measurements are unlikely to completely relieve the difference, as they probe the history of the Universe billions of years apart from each other and they would have to skew the results in the opposite directions.
One of the possibilities is decaying dark matter (DCDM) where dark matter particle decays partly into dark radiation. As radiation redshifts faster than dark matter, it results in reduced expansion rate at late times as compared to the early times.
Therefore, in DCDM model, the Hubble parameter at $z=0$, $H_0$, can be put in agreement with the evolution of $H(z)$ at higher redshifts, as measured in, {\it e.g.,} the CMB.

From the point of view of the impact on cosmology the scenario under consideration has significant similarities with the DCDM model. Therefore, in this section we describe the details of the analysis for the latter and later we use the obtained results to constrain our model.

We used publicly available Boltzmann solver code \texttt{CLASS}~\cite{Blas_2011} in combination with MCMC code \texttt{MontePython}~\cite{Brinckmann:2018cvx,Audren:2012wb} to constrain DCDM model and compare with standard $\Lambda$CDM cosmology.

We use the following data, with likelihoods already implemented in latest release of \texttt{MontePython}:

\begin{itemize}
     \item Planck 2018 measurements of the CMB~\cite{Aghanim:2019ame} (\texttt{TTTEEE} high-$\ell$, \texttt{TT}, \texttt{EE} low-$\ell$ and lensing likelihoods),
    \item BAO data from the BOSS survey~\cite{Alam:2016hwk,Beutler_2011,Ross_2015}.
    \item The galaxy cluster counts from Planck catalogue (PC)~\cite{Ade:2015fva},
    \item The local measurement of the Hubble constant (HST), $H_0=74.03\pm 1.42\;\mathrm{km/s/Mpc}$~\cite{Riess:2019cxk}.
\end{itemize}

In addition to 6 standard cosmological parameters $\{\omega_b, \omega_{cdm}, \ln 10^{10} A_s, n_s, 100 \theta_s, \tau_{reio}\}$~\cite{Aghanim:2018eyx}, we scan over two additional ones: $\Gamma$ and $\text{F}$. They denote decay width and fraction of DCDM that decays into dark radiation, respectively. Note that in the context of our model, the latter parameter was already introduced in Eq.~\eqref{eq:F}, while $\Gamma$ is the total decay width of $S$. We use thus obtained cosmological limits on $\Gamma$ and $\text{F}$ to find the parameter space regions of our model that is preferred from the perspective of cosmological data.

\subsubsection{Cosmological scan}

We performed three separate scans using in each case the same likelihoods. They correspond to $\Lambda$CDM, DCDM with broad prior on decay lifetime (later called \textit{short}) and DCDM with prior on decay lifetime constrained to be comparable to current age of the Universe  (later called \textit{long}). The last scan is motivated by~\cite{Vattis:2019efj} which found late DCDM model with lifetime $\sim20\,$Gyr can relieve the  Hubble tension. In this last case we fixed the reionization time, initial perturbation amplitude $A_s$ and its spectral index $n_s$ to $\Lambda$CDM best fit value, similar to what was done in~\cite{Vattis:2019efj}.

We used flat priors for 6 $\Lambda$CDM parameters with ranges set as follows:
$\omega_b=\Omega_b h^2 \in (0.01,0.1)$, $\omega_{\rm cdm} \in (0.05,0.3)$,
$100\theta_{s} \in (0.8,1.2)$, $\tau \in (0.01,0.2)$, $\ln (10^{10}A_s) \in (2,4)$, $n_s \in (0.9,1.1)$.
For two additional parameters, we used the same prior for amount of dark radiation coming from decay: $\log_{10}{\text{F}} \in (-4,-0.4)$, while using two different priors on the lifetime of DCDM, corresponding to short and long regimes: $\log_{10}{\Gamma}~\in (2,7)\ \rm [km/s/Mpc]$ and $\log_{10}{\Gamma}~\in (0,3)\ \rm [km/s/Mpc]$, respectively.

\begin{figure}[t]
\centering
\includegraphics[scale=0.55]{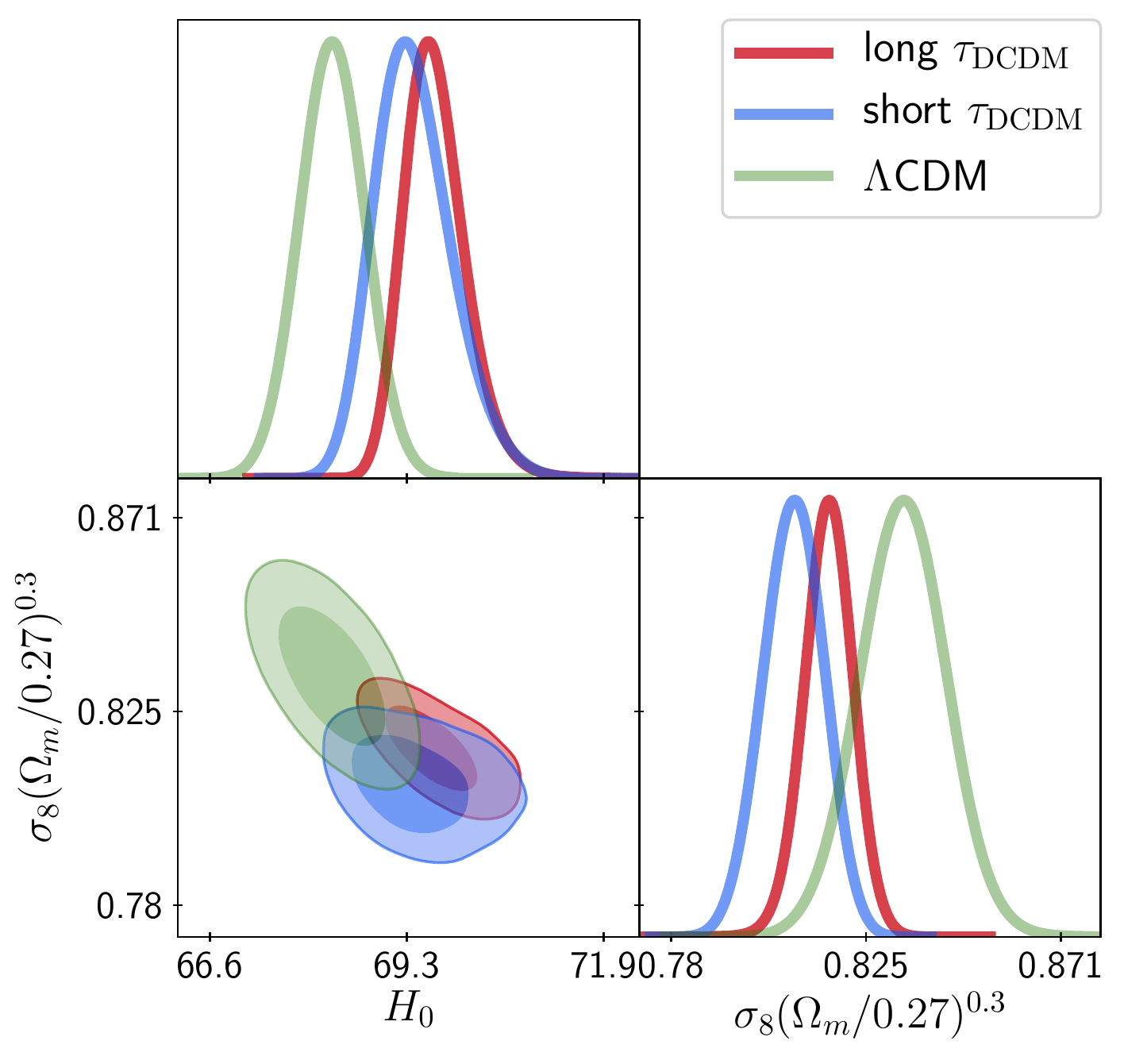}
\caption{Comparison of the fit for  $H_0$ and $\sigma_8$ in long and short decay lifetime DCDM models with the $\Lambda$CDM.}
\label{fig:cosmofit_compare}
\end{figure}

\begin{figure}[t]
\centering
\includegraphics[scale=0.32]{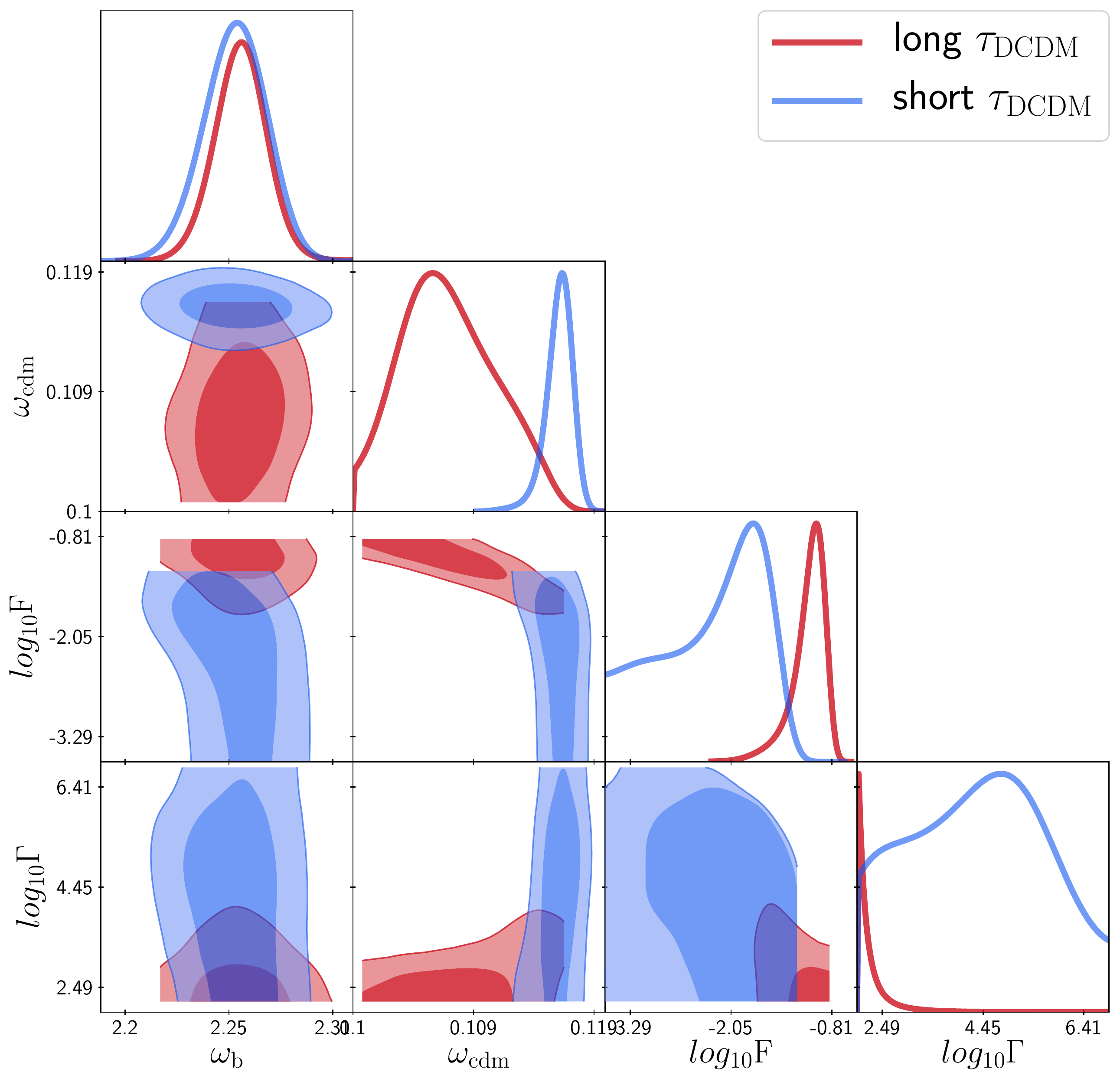}
\caption{Comparison of the two regimes of DCDM lifetime: long (red) and short (blue). Contours are given at the 1$\sigma$ (68\%) level.}
\label{fig:cosmofit_short}
\end{figure}

We generated chains until the Gelman-Rubin criterion $R-1<0.2$ is satisfied. The results of the scans are presented in Fig.~\ref{fig:cosmofit_compare} and~\ref{fig:cosmofit_short}.

We find two disconnected regions that improve the fit by mildly increasing $H_0$, relatively to $\Lambda$CDM. They correspond to early decay lifetime ($\sim 4\,$Myr) with small ($\sim$ 1\%) fraction going into dark radiation and to late decay lifetime ($\sim 5\,$Gyr) with significant fraction ($\sim$ 10\%) going into dark radiation. Such anticorrelation between F and $\Gamma$ is expected and was previously noted in {\it e.g.}~\cite{Poulin_2016,Chudaykin_2018}. In the first case,  all of $S$ decayed into $\chi$s by the onset of structure formation, therefore $\chi$'s self-interactions can improve the structure formation at small scales relative to the $\Lambda$CDM. In the second case, a potentially large fraction of final DM component is still in the noninteracting form of $S$ particles that did not yet managed to decay until the present day. In this case the scattering cross section can be even larger and therefore even tiny fraction of ultra-SIDM can serve as seeds of SMBHs.

Comparison with $\Lambda$CDM in $H_0$--$\sigma_8$ plane is shown in Fig.~\ref{fig:cosmofit_compare}. Mean values of the parameters are presented in Table~\ref{tab:bestfit_short}.
We see mild reduction in tension between CMB and low-redshift observations of $H_0$ and $\sigma_8$ in DCDM model.

\renewcommand{\arraystretch}{1.4}
\begin{table}
\centering
\begin{tabular}{|l|c|c|c|c|}
\cline{2-4}
\hline
      & short & long & $\Lambda$CDM  \\
\hline
100$\omega_{b}$      & $2.26_{-0.015}^{+0.017}$ & $2.26_{-0.014}^{+0.013}$ & $2.254_{-0.014}^{+0.014}$\\
$\omega_{cdm}$       &  $0.116_{-0.00084}^{+0.0011}$ &$0.107_{-0.0043}^{+0.0032}$ & $0.118_{-0.001}^{+0.001}$\\
$n_{s}$              &  $0.972_{-0.0040}^{+0.0043}$ & 0.9654   & $0.9705_{-0.0039}^{+0.0038}$\\
$10^{9}A_{s}$     &  $2.05_{-0.030}^{+0.032}$ & 2.106 & $2.107_{-0.037}^{+0.036}$  \\
100$\theta_{s}$      &  $1.04_{-0.00029}^{+0.00029}$  &  $1.04_{-0.00047}^{+0.00036}$ & $1.042_{-0.00029}^{+0.00029}$ \\
$\tau_{reio}$        & $0.0475_{-0.0070}^{+0.0080}$ & 0.0557 & $0.0578_{-0.0085}^{+0.0077}$ \\
\hline
$\log_{10}{\text{F}}$        & $-2.41_{-0.48}^{+0.96}$ &   $-1.1_{-0.081}^{+0.25}$  & - \\
$\log_{10}{\Gamma}$    & $4.36_{-1.49}^{+1.38}$ &    $2.33_{-0.33}^{+0.13}$ & - \\
\hline
$H_{0}$        & $69.4_{-0.60}^{+0.43}$ &  $69.7_{-0.44}^{+0.33}$ & $68.28_{-0.45}^{+0.45}$ \\
$\sigma_8$           & $0.791_{-0.0051}^{+0.0062}$  &  $0.80_{-0.0031}^{+0.0030}$ & $0.8065_{-0.0077}^{+0.0073}$ \\
\hline
\end{tabular}
\caption{Constraints on cosmological parameters. The uncertainties on the mean values are given at the 1$\sigma$ (68\%) level. The $\Gamma$ and $H_0$ are given in units of km/s/Mpc.
\label{tab:bestfit_short}}
\end{table}

\subsubsection{Structure formation}
\label{sec:structures}

Late time decays can affect not only the Hubble parameter, but also structure formation as the product of the decay can obtain sufficient energy to free-stream.

We impose the bounds coming from halo mass-concentration, galaxy-cluster mass function and Lyman-$\alpha$ power spectrum ~\cite{Peter:2010au,Wang:2013rha,Kaplinghat:2005sy,Borzumati:2008zz,Aoyama:2014tga} as an upper bound on mass splitting
between decaying (mother) particle and the resulting massive (daughter) particle. It provides the so-called kick velocity to the daughter particle, which at time of decay is $v_{\mathrm{kick}} \sim \delta$.

We can estimate the free-streaming length of daughter particle using formula from~\cite{Aoyama:2014tga}:
\begin{eqnarray}
\lambda_{\mathrm{fs}}&=\int_{\tau_{\mathrm{d}}}^{\tau_{0}} d \tau v(\tau) \sim \frac{3 v_{\mathrm{kick}}\Gamma^{-1}}{a_{\mathrm{d}}},
\label{eq:free_streaming}
\end{eqnarray}
where here $\tau$ is the conformal time, integration limits are conformal times corresponding to the time of decay and to the present, $\Gamma$ is the decay width and $a_d$ is the scale factor at the time of decay.

Lifetimes considered herein, correspond to $\Gamma^{-1}\lesssim 10\,$Gyr for which mass splitting is constrained~\cite{Kaplinghat:2005sy,Borzumati:2008zz,Aoyama:2014tga} to be: $\delta \lesssim 10^{-2}$ for short lifetime regime and $\delta \lesssim 10^{-3.5}$ for long lifetime regime (note Fig.~11 of~\cite{Aoyama:2014tga}). It is worth noting that in short lifetime regime, virtually all of $S$ will decay into self-interacting DM, and the elastic scatterings between the DM particles additionally should suppress free-streaming and somewhat relax the bound on mass splitting. For longer lifetime regime, the limits are stronger because the daughter particle had less time to redshift.

\section{Results}
\label{sec:results}

In this section we present and discuss the results of numerical scans for the three regimes A, B and C. In all the cases we implicitly assume that the correct observed relic abundance of DM is set by adjusting the details of the freeze-out and decay process of $S$.

\subsection{The SIDM regime}

\begin{figure*}[t]
\begin{subfigure}{0.49\textwidth}
\centering
\includegraphics[scale=0.5]{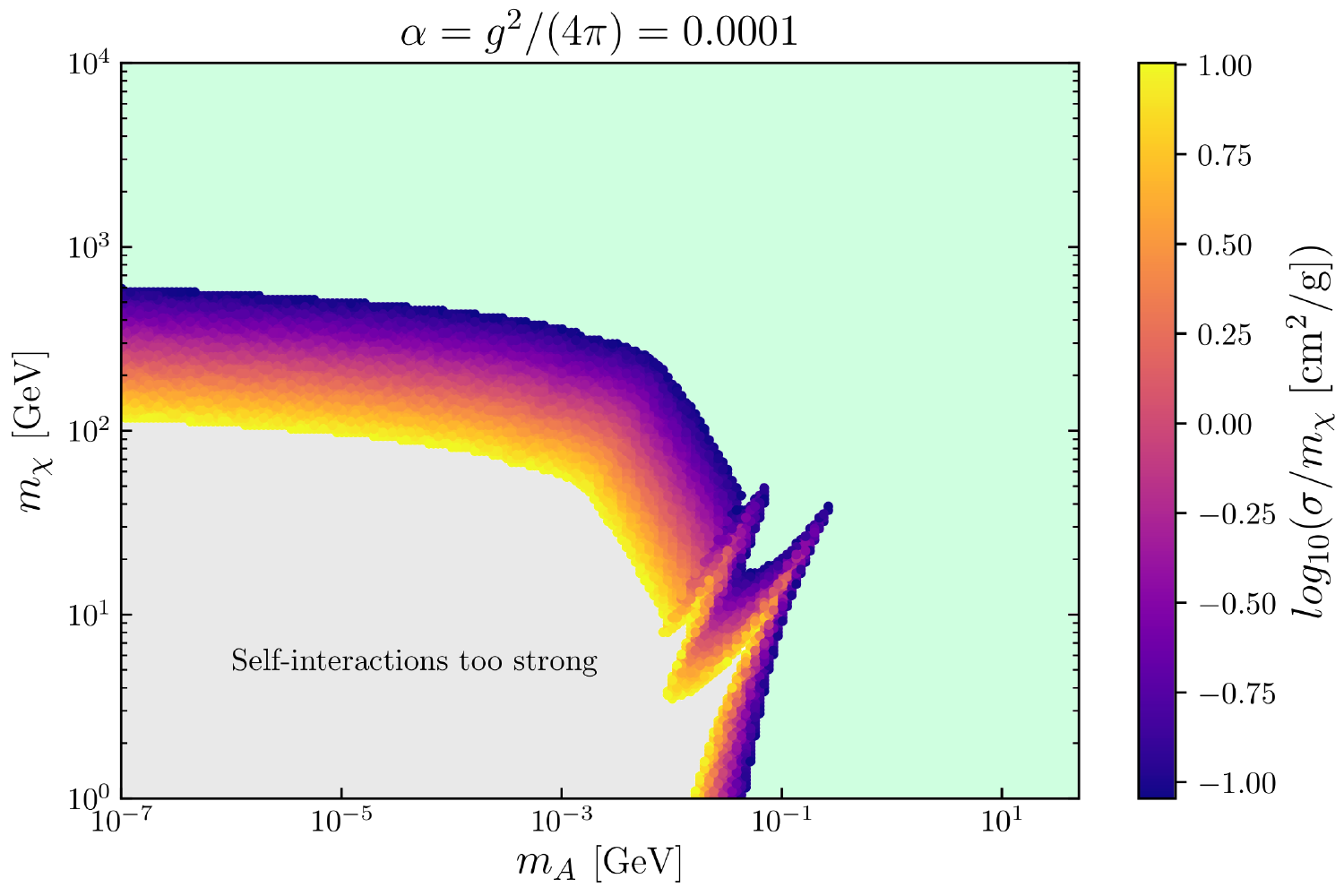}
\end{subfigure}
\begin{subfigure}{0.49\textwidth}
\centering
\includegraphics[scale=0.5]{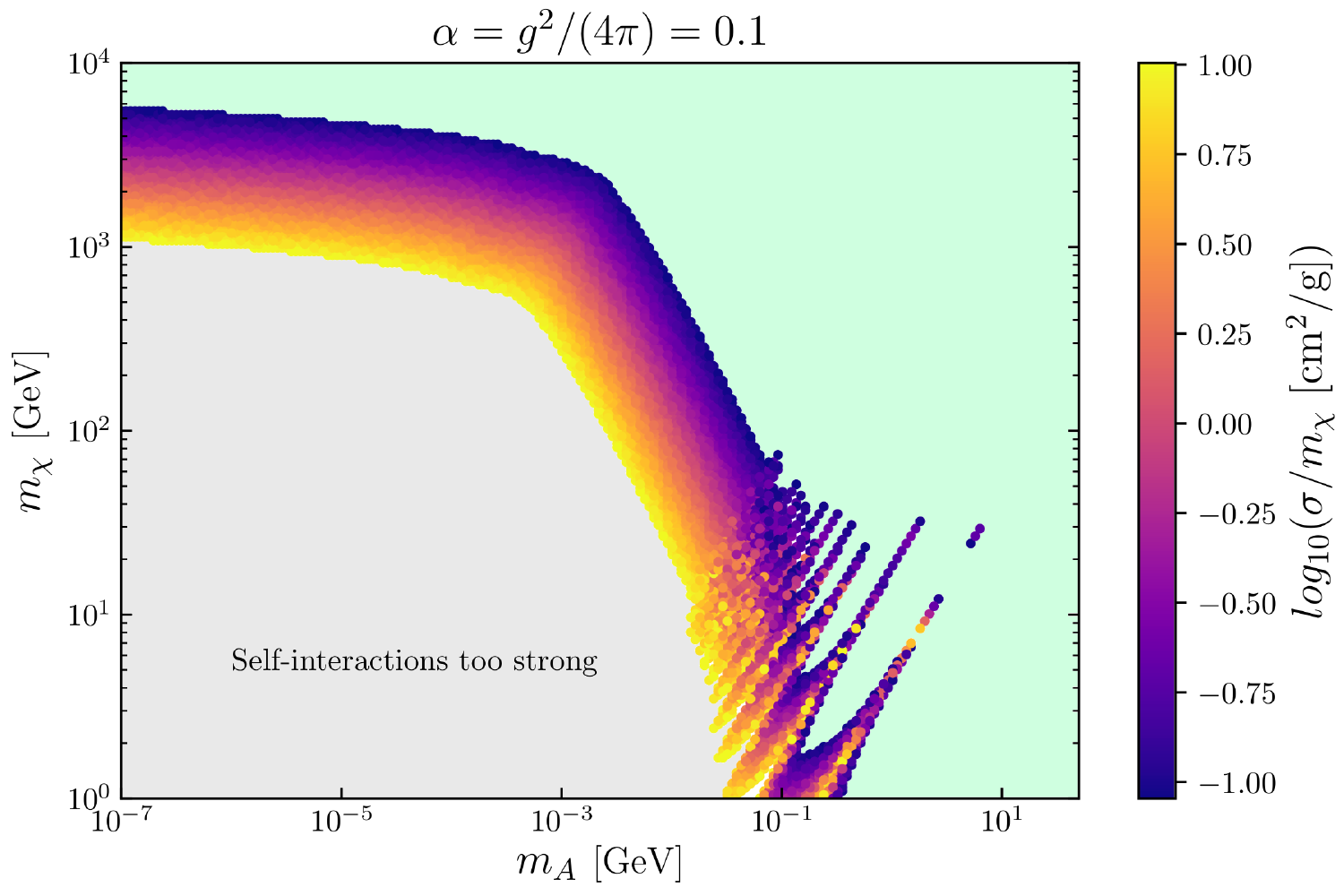}
\end{subfigure}
\caption{Regime A results for $\sigma/m_\chi$ in the range 0.1--10 cm$^2$/g preferred by the SIDM solution of the small scale problems in the $m_A$--$m_\chi$ plane for two representative values of coupling constant $\alpha =0.0001$ (left) and $\alpha=0.1$ (right). The gray area on the bottom left is excluded as it leads to too strong DM self-interactions, while the pale green region above is allowed, but does not affect structures at small scales.
}
\label{fig:results_1}
\end{figure*}

For the $\epsilon$ values small enough that the dark sector does not thermalize with the SM, but at the same time large enough that $S$ decays happen before recombination the scenario effectively boils down to  a self-interacting $\Lambda$CDM model. Phenomenologically it has the same properties as many well studied SIDM models (again we refer to, {\it e.g.,}~\cite{Tulin:2017ara} for a review), with two important distinctions. First, the self-interaction strength is governed by a different coupling that the one giving rise to the relic abundance, opening much wider parameter space. And second, the light mediator can be completely stable rendering the most constraining limits ineffective. In this regime the whole phenomenology is governed by $m_\chi$, $m_A$ and $\alpha$.

In Fig.~\ref{fig:results_1} we present the cross sections of the strength needed for solving small-scale structure problems of $\Lambda$CDM with rainbowlike palette. The left panel shows the case of small $(\alpha=10^{-4})$ while the right panel large $(\alpha=10^{-1})$  values of the coupling. One can notice well-known resonant behavior in lower right part of the plot, which gets more pronounced as $\alpha$ increases. For fixed $m_A$, correct $\sigma/m_\chi$ is inversely proportional to $m_\chi$ and directly proportional to $\alpha$, as expected. In gray region, parameter space is excluded due to too strong self-interactions~\cite{Markevitch:2003at,Clowe:2006eq,Randall:2007ph}.
The light-green region predicts too weak self-interactions to affect cosmology at the small scales in any visible way. The existence of color bands in between, spanning more than an order of magnitude in both masses when taking into account varying $\alpha$ is a demonstration that the proposed mechanism can successfully give rise to the viable SIDM candidate.

Before ending this section let us mention that even in the regime where the $S$ decays happen well before recombination the resulting DM component can help alleviate the cosmological tensions. This was observed and studied in detail in \cite{Bringmann:2018jpr} where it was found that if annihilation happens very close to the peak of one of the Sommerfeld effect resonances, the DM can undergo a second period of annihilations at late times \cite{vandenAarssen:2012ag}, leading to conversion of some fraction of matter to radiation. The same effect can appear in our setup, with the modification due to different thermal histories of the DM component. In particular, in \cite{Bringmann:2018jpr} the time of kinetic decoupling from the SM thermal bath plays a significant role. However, if $\chi$s came from decays of $S$ and were never in equilibrium, then the evolution of their velocity distribution, and consequently the impact of possible late time Sommerfeld enhanced annihilations, would require a separate study.

\subsection{The SIDM from late decays regime}

\begin{figure}[t]
\centering
\includegraphics[scale=0.52]{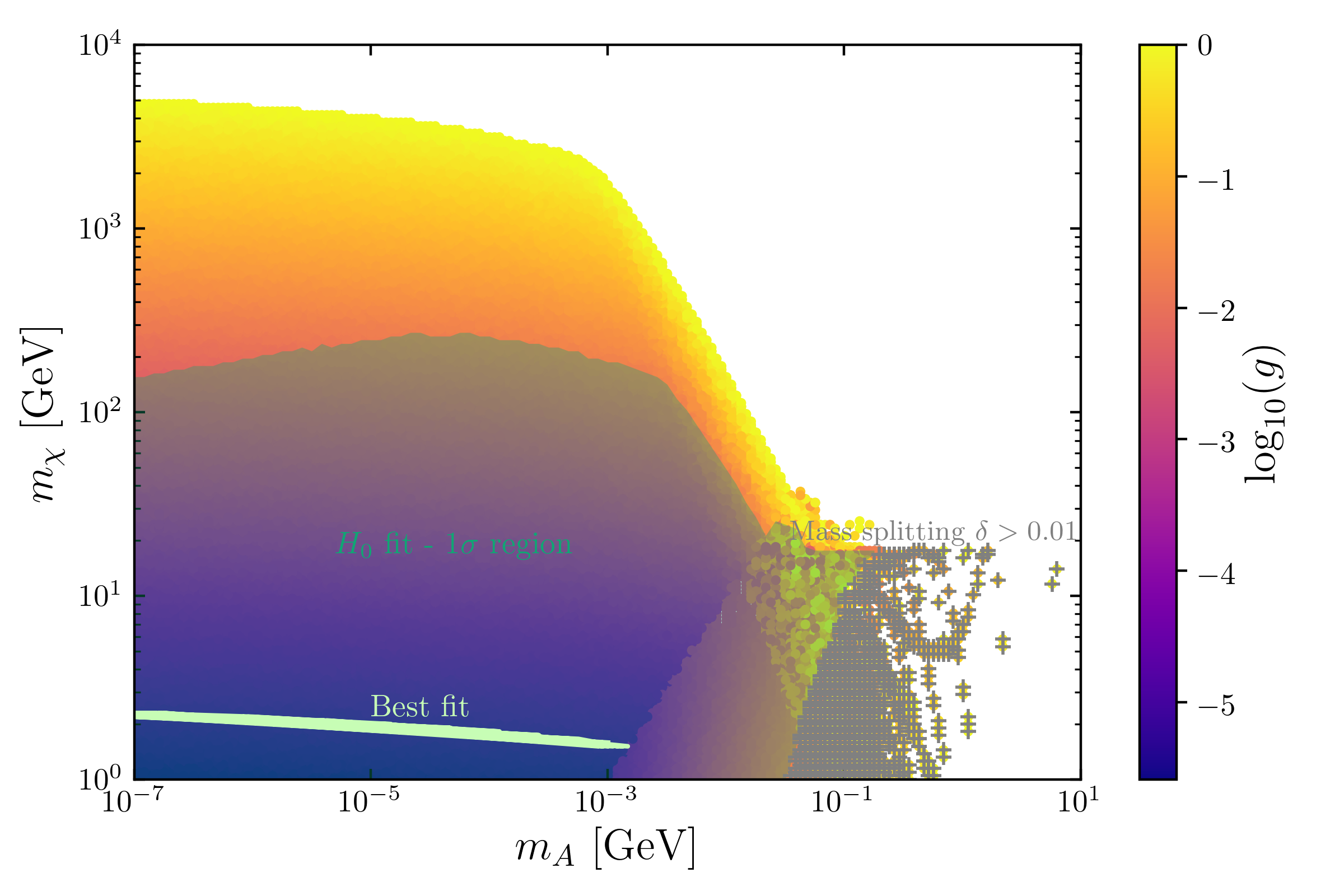}
\caption{The results for the SIDM regime B originating from late $S$ decays. Color coding denotes the value of the coupling $g$ for the points that satisfy the condition $\sigma/m_\chi \sim (1\pm10\%)$ cm$^2$/g. On top of that the dark green shade denotes the region at the 1$\sigma$ (68\%) level around the mean values of DCDM parameters, which relax Hubble tension in the short lifetime scenario. Gray pluses overlay points that have $\delta>0.01$ which are in this model in tension with the structure formation.}
\label{fig:results_2}
\end{figure}

Lowering the $\epsilon$ values, the lifetime of $S$ extends beyond the recombination and the following decays modify the cosmological model. In this regime the resulting dark matter phenomenology is still governed by $m_\chi$, $m_A$ and $\alpha$, but $m_S$ (or equivalently $\delta$) and $\epsilon$ start to have important consequences as well by affecting the kinematics of the decay and the lifetime, respectively.

The main results for this regime are given in Fig.~\ref{fig:results_2}. It shows the results of the cosmological scan with priors set to short $S$ lifetime projected onto fixed\footnote{The $\sigma/m_\chi$ was fixed to a representative value in order to enhance readability of this particular figure, while we emphasize that allowing larger range for the cross section enlarges the allowed parameter space.} $\sigma/m_\chi \sim (1\pm10\%)$ cm$^2$/g in the $m_A-m_\chi$ plane, with colour bar indicating coupling strength $g$.
The dark semitransparent green region shows 1$\sigma$ range around the
best fit values relaxing the Hubble tension, {\it i.e.} the DR fraction of $F=10^{-2.41}\approx 0.004$. The light green line denotes the best fit parameters, which depend on $m_S$, hence it is a continuum and not a point, with small width due to numerical resolution.

The numerical scan was performed in a grid over four parameters uniquely specifying this fraction: $m_S$, $m_A$, $m_\chi$ and $g$, with the condition that the mass splitting, Eq.~\eqref{eq:mass_splitting}, is small, $\delta \in [10^{-6}, 10^{-1}]$.
The only remaining relevant cosmological parameter, the decay width $\Gamma$, can always be brought to correct value by rescaling the $\epsilon$ coupling constant.

 The lower right region starting roughly at the right tip of best fit and going along right diagonal, represents the resonant regime. One sees smaller density of points here, compared to Born and classical regimes, and higher values of $g$ are allowed. For largest $m_A$, points are very sparse which comes from the irregular pattern of consecutive resonances which have very small width for large value of $\alpha$. Roughly half of resonant parameter space is also marked by gray pluses, which denote that those points require large $\delta$, which is in tension with structure formation limits.

The $1\sigma$ region is bounded from above by the condition on $F$. The points above this bound are giving too efficient conversion to DR and manifest in two regimes. The resonant and $\alpha\sim 1$ regimes are dominated by loop decay into two $A$s. This region is, partially, also constrained by the limit on $\delta$. For the rest of the parameter space, three body decay of $S$ is dominant.

It is worth stressing that a large parameter space of the model allows for both the self-interactions to be at the right range to potentially solve small scale problems and to decay to correct amount of radiation to help relieving the $H_0$ tension.

\subsection{The uSIDM regime}
\label{sec:uSIDM}

Finally, for even longer $S$ lifetimes we enter the two-component DM regime where the $\chi$ can be much more strongly interacting.
As was noticed in~\cite{Pollack:2014rja} and followed by, {\it e.g.,}~\cite{Choquette:2018lvq}, such uSIDM could provide a mechanism of formation of supermassive black holes with masses of order $10^9 M_{\rm Sun}$ which formed by $z\sim 7$. Such SMBHs were observed recently~\cite{Mortlock_2011,DeRosa:2013iia,Banados:2017unc} and provide a challenge for standard formation mechanisms because of their large masses forming at such an early time. The proposed mechanism of~\cite{Pollack:2014rja} is similar to ordinary gravothermal collapse which is believed to be responsible for formation of globular clusters~\cite{LyndenBell:1968yw} and takes place by ejection of most energetic stars, allowing the rest of the system to contract. Concerning black holes formation, uSIDM causes similar process in DM halo and as there is no inhibitor to the process, SMBH forms. Unfortunately, if uSIDM constitutes the whole of DM, self-interaction rate necessary for gravothermal collapse exceeds the bound set by, {\it e.g.,} the Bullet cluster. However, a small fraction of even ultra strongly interacting DM is allowed by observations and as showed by detailed simulations in a framework of multi-component DM models in~\cite{Pollack:2014rja,Choquette:2018lvq}, can be responsible for boosting the formation rate of SMBHs.

\begin{figure}[t]
\centering
\includegraphics[scale=0.55]{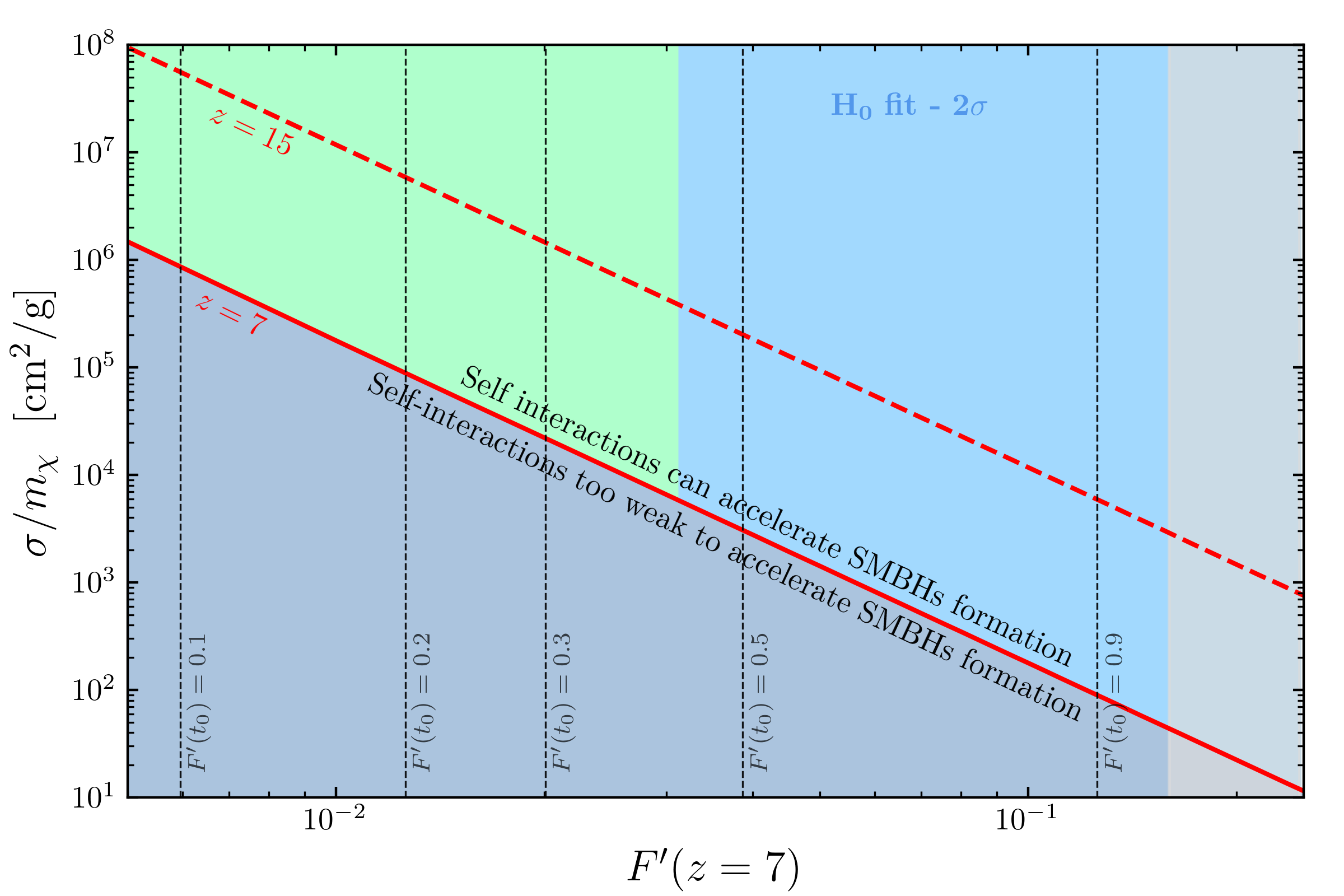}
\caption{The results for the regime C. The blue and green regions feature self-interactions strong enough to accelerate SMBHs formation rates, while on top of that the blue region is in the $2\sigma$ region around the best fit for the $H_0$ parameter. The dotted vertical lines show contours of uSIDM fraction at the present day. See the text for more details.}
\label{fig:results_3}
\end{figure}

In Fig.~\ref{fig:results_3} we show the results of the long lifetime scan in the $F'$--$\sigma/m_\chi$ plane, where $F'$ denotes the fraction of uSIDM that existed by $z=7$. In light blue we show a region at the 2$\sigma$ (95\%) level around the mean values of DCDM parameters which relax  Hubble tension in the long lifetime scenario. Vertical dashed lines denote resulting fractions of uSIDM component at the present day. These are significantly larger than the values of $F'$ on the $x$-axis, because of the decays that take place between $z=7$ and $z=0$. It follows that the whole light blue region leads to a scenario where ultra strongly interacting component constitutes unacceptably large fraction ($\gtrsim 0.4$) of DM at late times. Therefore, we find that if the uSIDM  arises from decays of an intermediate unstable state the requirement of significant fraction of uSIDM to be already present at $z\sim 7$ implies very long lifetimes $\gtrsim 40$~Gyr giving small fraction $F'$ and large scattering cross sections. This is not the parameter region that is preferred for the requirement of relaxing the Hubble tension.

The light green region denotes the parameter space where decay of $S$ happens too late to significantly influence the $H_0$ tension, but with large enough $\sigma/m_\chi$ and $F'$ to be relevant for accelerating SMBHs formation. Therefore, significant part of the parameter space corresponds to a scenario of two-component DM which provides a viable mechanism of production of subdominant uSIDM. Note that although some parts of this region  lead to a substantial present day uSIDM component as well, the exact limits on $F'(t_0)$ are rather uncertain and do not exclude the whole parameter space of the model.

The red lines are the results of numerical simulations performed in~\cite{Choquette:2018lvq} (Fig.~5, Model A for elastic scatterings) and denote redshifts $z=7$ (solid) and $z=15$ (dashed).
In that work, two component DM scenario was assumed, with constant fraction of uSIDM, $F'$. In our case, $F'$ depends both on time of the decay $1/\Gamma$ and the fraction $1-F$ going into DM component. Hence, the limits presented here should be taken as exemplary and further study conducting numerical simulation would be needed.

To summarize, we find that production of uSIDM via late decay is strongly constrained if one restricts the decay lifetime to be $\lesssim 40\,$Gyr, which would at the same time relax the Hubble tension. The difficulty lies at the very early time of SMBH formation as $z\sim7$ corresponds to $\sim0.77\,$Gyr, while we find that the decay times relevant to Hubble tension correspond to either earlier ($\sim 4\,$Myr) or longer ($\sim 5\,$Gyr) times. However, if the decay is assumed to happen even later than $40\,$Gyr, it could be a viable mechanism of accelerating the formation of early SMBHs.

\section{Discussion and conclusions}
\label{sec:conclusions}

Motivated by the question of how far in solving or alleviating the tensions of $\Lambda$CDM one can go by modification of only the dark matter component in a complete particle physics model, we study in this paper the implications of the self-interacting dark matter production mechanism based on (late time) decays of an intermediate thermally produced WIMP-like state. The decay is at the tree-level only into pair of DM particles, while at higher order three-body and loop processes introduce small branching ratio to final states containing the light mediator. This leads to a very natural explanation of why only several percent of the dark matter energy was transferred into radiation, which is a necessary condition for improving the fit to the $H_0$ parameter. At the same time, if only the lifetime of the intermediate state is smaller than the age of the Universe, the whole noninteracting dark matter is converted into strongly interacting component capable of addressing as well the $\Lambda$CDM tensions at small scales. Moreover, this mechanism allows the mediator to be stable and therefore avoid strong limits from the observations of cosmic microwave background and indirect detection.

From a particle physics perspective such scenario is a natural extension of the very well studied models connecting the dark sector with the visible sector by a weak portal. We provide and study a simple example model of this kind, where for concreteness we focus on the Higgs portal. Within this model we perform numerical analysis with the emphasis on the  dark matter self-interaction properties and fits to local and global cosmological measurements. We find that the proposed mechanism allows for a perfectly viable self-interacting dark matter with large parameter space resulting in the elastic cross section of the correct range to address the small scale cosmological problems. Additionally, there exists a significant overlap with the parameter regions required to impact the Hubble tension. However, the resulting improvement of the fit is relatively mild, not offering any improvement over alternative methods to reduce the $H_0$ tension.

We also consider a scenario when the decays happen much later, with lifetimes of order $\mathcal{O}(1\,\rm Gyr)$ or larger. We find that in that regime the resulting dark matter consists of two components: dominant noninteracting one and a subdominant component of SIDM or uSIDM type. Although the former has typically too small number density to address the small scale problems of $\Lambda$CDM in that scenario, the latter provides a viable model of ultra strongly interacting DM that can help accelerate the formation of the SMBHs and by doing that explain how could they have been formed at times as early as $z\sim 7$. Unfortunately, within the studied model we find that the region that could simultaneously alleviate the Hubble tension and provide a mechanism for speeding up the SMBHs formation is not allowed by the observations due to unacceptably large uSIDM component at the times of the Bullet cluster at $z\sim 0.5$.

It is worth adding that although all the explicit results given in this work are for the DM being a Dirac fermion interacting via a vector mediator, we have also analyzed a scenario in which the mediator is a scalar leading to purely attractive interactions. This does not introduce any qualitative change and also quantitatively the results are  similar to those presented in Fig.~\ref{fig:results_2}.

Last but not least, let us comment on the recently reported  unaccounted excess of events over the background in electronic recoils around 1--7 keV in the XENON1T experiment~\cite{Aprile:2020tmw}. One of the potentially most promising explanations of this excess in terms of new physics involves the existence of a dark photon coupled to SM via kinetic mixing term $-\frac\kappa 2 F_{\mu\nu}F^{\prime\mu\nu}$~\cite{Alonso-Alvarez:2020cdv}. What was noticed in that paper, is that both XENON1T excess and observations of cooling anomalies in horizontal branch stars~\cite{Raffelt:1987yu,Ayala:2014pea,Giannotti:2015kwo} could be explained by light $\sim$ keV dark photon with kinetic mixing parameter $\kappa\sim 10^{-15}$.

It is interesting to note that the light mediator $A^\mu$ studied in this work for completely independent reasons, is in fact also the same as the aforementioned dark photon. Although, in our work for simplicity we considered no kinetic mixing in the interaction Lagrangian, one can naturally incorporate it as long as $\kappa\lesssim 10^{-12}$, {\it i.e.,} when the resulting interactions will not significantly affect the thermal history of neither the DM nor the light mediator.\footnote{Allowing kinetic mixing dark photon becomes unstable, however, with such low mass it can only decay to three photons which leads to lifetimes much greater than the age of the Universe~\cite{Pospelov:2008jk,Alonso-Alvarez:2020cdv}.} Therefore, it is intriguing to note that allowing $\kappa\sim 10^{-15}$ could be relevant to XENON1T excess, in addition to production of SIDM while simultaneously mildly relaxing the Hubble tension. This could serve as a compelling motivation to perform further dedicated studies of models featuring the production mechanism put forward in this work.

\acknowledgments
We would like to thank Torsten Bringmann and Leszek Roszkowski for valuable comments. A.H. is supported in part by the National Science Centre, Poland, research grant No. 2018/31/D/ST2/00813. K.J. is supported in part by the National Science Centre, Poland, research grant No. 2015/18/A/ST2/00748. The use of the CIS computer cluster at the National Centre for Nuclear Research in Warsaw is gratefully acknowledged.

\bibliography{biblio}

\end{document}